# Materials with ZrCuSiAs-Type Structure

Rainer Pöttgen[a], Dirk Johrendt[b]

[a] Institut für Anorganische und Analytische Chemie, Universität Münster, Corrensstrasse 30, 48149 Münster, Germany

[b] Department Chemie und Biochemie, Ludwig-Maximilians-Universität München, Butenandtstrasse 5-13 (Haus D), 81377 München, Germany

Reprint requests to R. Pöttgen. E-mail: pottgen@uni-muenster.de

The discovery of high-temperature superconductivity in the fluoride doped arsenide oxides *RE*FeAs($O_{1-x}F_x$) (*RE* = early rare earth element) with transition temperatures as high as 55 K gave a true renaissance in superconductivity research. These arsenide oxides constitute a rather small fraction of much larger family of compounds with the tetragonal ZrCuSiAs type structure (space group *P*4/*nmm*), among them pnictide-oxides and -fluorides, chalcogenide-oxides and -flourides as well as silicide and germanide hydrides. Besides the spectacular superconductivity, these materials have further interesting properties with respect to magnetic ordering, transparent semiconducting behavior or to optical properties. The crystal chemical and chemical bonding peculiarities as well as the broadly varying physical properties are reviewed herein.

*Key words*: Pnictide Oxides, Chalcogenides, Superconductivity, Semiconductors

## Introduction

The quaternary silicide arsenides ZrCuSiAs and HfCuSiAs have been reported in 1974 by Johnson and Jeitschko [1]. This structure type is tetragonal, space group *P*4/*nmm* with two formula units per cell. A similar atomic arrangement has been observed for the ternary silicide $HfCuSi_2$ [2]. These compounds can be considered as filled-up variants of the PbFCl [3-5] type structure. The latter derives from the $Cu_2Sb$ type. More than 260 binary and ternary intermetallic compounds with this structure are listed in the Pearson Handbook [6]. Compounds with $Cu_2Sb$ type structure show different coordination and chemical bonding patterns depending on the size or valence of the constituting elements and have been reviewed by Pearson [7].

Further representatives of the ZrCuSiAs type were reported in 1980 by Palazzi *et al.* [8-10]. They reported the sulphide oxides LaCuSO and LaAgSO and investigated the silver ion conductivity. Afterwarrds, many other sulphide and selenide oxides have been reported. This family of compounds has attracted considerable interest in recent years since Ueda and coworkers reported that LaCuSO has an optical band gap of 3.1 eV and exhibits *p*-type electrical conductivity [11, 12]. These materials have potential application as *p*-type transparent semiconductors.

In parallel to the sulphide oxides, several rare earth (*RE*) based pnictide (*Pn*) oxides have been reported by the group of Jeitschko [13-20]. These compounds were first believed to be ternary pnictides [21], however, careful synthesis yielded pure pnictide oxides. These pnictide oxides can be described with the electron precise formulæ $RE^{3+}T^{2+}Pn^{3-}O^{2-}$ ($T$ = late transition metal). Indeed, several of these compounds are transparent with yellow to dark red color and they have been studied with respect to their optical properties [22, 23]. LaFePO [24, 25] and LaNiPO [26, 27] show superconductivity below 3.2 and 4.3 K, respectively.

Outstanding in this family of compounds are the arsenide oxides which can be doped with fluoride, leading to solid solutions $RETAsO_{1-x}F_x$. This doping induces transitions to superconductivity at comparatively high temperatures, *e. g.* $T_C$ = 26 K in $LaFeAsO_{1-x}F_x$ ($x$ = 0.05–0.12) [28]. Most recently an even higher transition temperature of 52 K has been reported for the corresponding praseodymium system [29]. Since that discovery in March 2008, numerous new contributions are published daily on the preprint server of the Cornell University Library (arXiv.org). A short highlight on the recent developments has been published recently [30].

So far more than 150 representatives have been reported for the family of ZrCuSiAs type compounds (Table 1). The latter display a broad crystal chemical variety and different patterns in chemical bonding. The compounds have outstanding physical properties in the fields of magnetism, superconductivity, and transparent semiconductors and a large potential for applications. The crystal chemical details and the recent developments regarding the physical properties are reviewed herein.

## Synthesis

The compounds listed in Table 1 have been synthesized via different techniques, depending on whether polycrystalline samples or small single crystals are required.

The phosphide oxides have first been obtained during synthesis of ternary rare earth (*RE*)–transition metal (*T*)–phoshides using the tin flux technique [21, 31], most likely due to surface contamination of the rare earth filings. Starting materials for these reactions were the elements using a large excess of tin as flux medium and silica tubes as crucible material. Since several of the phosphide oxides have remarkable stability, the tin matrix can be dissolved in moderately diluted (1:1) hydrochloric acid. For CeRuPO crystals with an edge length of 2 mm were obtained this way [32].

Later, equimolar NaCl/KCl mixtures were used as flux medium for crystal growth. In a typical experiment [22], filings of the rare earth element, powder of the transition metal oxide (*e. g.* MnO or ZnO) and powder of red phosphorous were sealed in evacuated silica tubes together with a ca. fourfold weight excess of the flux medium. The temperature profile for the annealing procedure depends on the starting components. In some cases also intermetallic precursor compounds can be used. β-PrZnPO was prepared from $PrZn_2$, $Pr_6O_{11}$, ZnO and P in an atomic ratio 11:2:1:23 under salt flux conditions [22]. The salt flux can easily be dissolved in demineralized water. It is thus a good alternative to the tin flux, since no acidic medium is needed for flux dissolution.

Mostly the salt flux technique results in smaller quantities of well shaped single crystals. For larger amounts of polycrystalline samples a ceramic route can be applied. The rare earth monopnictide (prepared by arc-melting or also *via* a ceramic route) is then mixed with the transition metal monoxide and pelletized. The pellets are annealed in evacuated sealed silica tubes up to 1370 K (depending on the constituents). Sometimes regrinding and additional annealing procedures are necessary in order to obtain sngle phase products. Using this route, e. g. LaMnPO [33] and the series *RE*MnSbO and *RE*ZnSbO [34] have been obtained.

Various doping experiments for investigating the superconducting behaviour with higher- or lower-valent metal cations have been introduced. Typical sources are zirconium metal or $Ca_3P_2$. The doping with fluoride can be performed via $FeF_3$ or the rare earth fluorides. Besides the transition metal oxides also the rare earth metal oxides can be used as oxygen source.

The pnictide fluorides have been synthesized with the difluorides of the alkaline earth (*AE*) element as a fluoride source according to the general equation $AEF_2 + AE + 2T + 2Pn \rightarrow 2AETPnF$. Sealed evacuated silica tubes had been used as crucible material and a special annealing sequence with a maximum temperature of 1170 K [35] had been applied. The reaction mixtures had first been heated at 770 K in order to account for the vapor pressure of the pnictogen in the initial reaction step. BaZnPF can also be synthesized *via* a metathesis reaction according to NaZnP + BaFCl → BaZnPF + NaCl [36]. The pnictide fluorides are stable in air.

The ZrCuSiAs type hydrides have been synthesized starting from the ternary intermetallic compounds. The latter can be obtained via arc-melting and subsequent annealing. As an example we give the synthesis conditions for CeCoSiH [37]. A CeCoSi ingot is first annealed under vacuum at 523 K for 12 h and then exposed to 4 MPa hydrogen gas at the same temperature. The hydrogenation procedure always induced a decrepitation of the starting ingot into small grains and the resulting hydrides are stable in air.

The chalcogenide oxides have mostly been prepared by classical solid state reactions. Starting materials were rare earth metals, rare earth sesquioxides, powders of the chalcogen, and copper powder [38]. The educts were all thoroughly mixed and ground to powders, pressed into pellets and annealed in evacuated silica tubes. The maximum annealing temperature depends on the chalgogen component. Longer reaction times up to 15 days and intermediate grindings may be necessary. In some cases, also freshly prepared copper(II)oxide (prepared by heating $Cu(NO_3)_2·2H_2O$ in air) can be used as an oxygen source [39]. Alternatively, also mixtures of $La_2S_3$, $Cu_2S$, and $La_2O_3$ can be used with alumina as crucible material. SrS is a suitable precursor for the synthesis of doped samples like $La_{1-x}Sr_xCuSO$. Thin films of LaCuSO and related sulphide oxides can effectively be prepared by radiofrequency sputtering [11]. Crystal growth of the CeAgSO [40] and *RE*CuTeO (*RE* = La, Ce, Nd) [41] compounds can be ameliorated using small amounts of KCl and KI as flux reagents. Crystals of LaCuSO were obtained by heating polycrystalline powder in an iodine atmosphere at 1373 K [42].

Besides the classical solid state reactions, the sulphide oxides are also accessible through hydrothermal reactions at significantly lower temperatures [43]. BiCuSO was obtained from $Bi_2O_3$, $Cu_2O$, and $Na_2S$ in a PTFE Teflon-lined pressure vessel using water as solvent. In contrast to the solid state reaction, the maximum temperature for the hydrothermal reaction was only 520 K. Another approach for the synthesis of BiCuSO is the use of reactive precursor compounds. Hiramatsu *et al.* [44] used freshly prepared $BiCu_3S_3$ (obtained from $Bi_2S_3$ and $Cu_2S$ under a flow of a 5% $H_2S$ and 95 % $H_2$ gas mixture at 570 K) and $\alpha$-$Bi_2O_3$ for the solid state reaction.

Platinum is a useful container material for the synthesis of fluoride sulphides. EuCuSF [45] was obtained from Eu, $EuF_3$, Cu and S in 2:1:3:3 ratios in a NaCl flux in a sealed platinum ampoule at 1120 K. For the synthesis of the BaCu*Q*F samples (*Q* = S, Se), CuS, CuSe, BaS, BaSe, Cu, and $BaF_2$ were used as starting reagents for the solid state reactions [38] at 720 K.

**Crystal chemistry**

The ZrCuSiAs type structure has more than 150 representatives. In Table 1 we have listed the basic crystallographic data of these compounds classified according to their *p*-element components [1, 8-10, 13-29, 32-88]. As emphasized in Figure 1, these compounds are composed of two different layers, both with tetrahedral coordination. The transition metal atoms have tetrahedral coordination by a heavier chalcogen or pnicogen atom, while the silicon, oxygen, and fluorine atoms fill the tetrahedral voids left by the electropositive elements. The different tetrahedral layers are stacked in an AB AB sequence. As emphasized below, chemical bonding within the layers has predominantly covalent character, while the interlayer bonding is of an ionic type.

Repeatedly, these materials have been classified as layer compounds; however, they do not meet the requirement for a layer structure, *i. e.* a van-der-Waals gap with weak interlayer bonding (such layers can easily be disrupted mechanically leading to softness of the material) like in graphite or molybdenum sulphide.

Most of the ZrCuSiAs type compounds are stable over a wide range of temperatures. So far, dimorphism with a rhombohedral NdZnPO type high-temperature modification has only been observed for CeZnPO and PrZnPO [19, 22].

The oxides, fluorides, and the recently reported hydrides in this family of compounds can be described, to a first approximation, with simple ionic formula splittings, e. g. $Pr^{3+}Zn^{2+}P^{3-}O^{2-}$, $La^{3+}Ag^{+}S^{2-}O^{2-}$ or $Ce^{3+}Fe^{2+}Si^{4-}H^{-}$. This description is certainly justified, since several phosphides and sulphides are transparent, while some arsenides, antimonides and the hydrides are black. In all cases, the rare earth containing layers are positively, and the transition metal containing layers nega-tively charged according to $[Pr^{3+}O^{2-}]^{+}[Zn^{2+}P^{3-}]^{-}$, $[La^{3+}O^{2-}]^{+}[Ag^{+}S^{2-}]^{-}$, $[Th^{4+}O^{2-}]^{2+}[Ag^{+}P^{3-}]^{2-}$, or $[Ce^{3+}H^{-}]^{2+}[Fe^{2+}Si^{4-}]^{2-}$, enabling carrier flows between the layers.

Although a precise formula splitting is possible for the ZrCuSiAs type compounds, several examples show significant defects on the transition metal site. If such defects occur, the systems switch to metallicity. Prominent examples are $BiCu_{0.97}SO$, $BiCu_{0.97}SeO$, $LaCu_{0.96}SO$, $LaCu_{0.94}SeO$, $LaCu_{0.96}TeO$ [44] and the series $CeCu_{1-x}SO$ [40], $CeCu_{1-x}SeO$ [39], and $CeAg_{1-x}SO$ [40]. In the latter case, the different formation of defects has a drastic influence on the magnetic behavior (*vide infra*).

An important parameter for the physical properties is the interlayer distance between the two layer types. Hiramatsu *et al.* [44] have studied the crystal structures of the two series LaCu*Ch*O and BiCu*Ch*O (*Ch* = S, Se, Te) in order to get reliable values for the interatomic distances. The *c* parameters strongly increase from 856 (BiCuSO) to 952 pm (BiCuTeO) and from 852 (LaCuSO) to 934 pm (LaCuTeO). Consequently one observes significantly different distances and bond angles within the square prisms around the lanthanum (bismuth) atoms and within the $CuCh_4$ tetrahedra.

## Chemical bonding

The two-dimensional nature of the ZrCuSiAs-type structure reflects the chemical bonding pattern. Bonding interactions between atoms of different electronegativities are responsible for a more or less pronounced ionic character of the ZrCuSiAs-type compounds. The bonding scenario can be subdivided in three main aspects: Firstly, all compounds are stabilized by electron transfer between the layers according to, for instance, $[LaO]^{+}[FeAs]^{-}$. Secondly, the bonds within the transition-metal layers [Fe–As] have more covalent character than those inside the oxide layers [La–O] and between the layers [La–As], which are rather ionic. This is in agreement with an analysis of the chemical bonding in the ZrCuSiAs-type hydride CeCoSiH, which showed the by far strongest covalent Co–Si bonds inside the [CoSi] layers [37]. The third point concerns direct metal-metal interactions between the transition metals, which may become significantly bonding if the *d*-shell is only partially filled.

The charge transfer between the layers is of course smaller than in the purely ionic picture. A Bader analysis of the charge distribution in LaOFeP resulted in $[LaO]^{0.36+}[FeP]^{0.36-}$ [89]. Electronic, optical and magnetic properties strongly depend on the character of the electronic states close to the Fermi level. In semiconducting close shell systems like *RE*ZnPO, the energy gap occurs between the P-3*p* valence band and the conduction band made of Zn-4*s*/P-3*p* antibonding states. The latter slightly hybridize with the 5*d*- and 4*f*-dorbitals of the *RE* atoms, which leads to subtle nuances of the gap and thus different colors of the compounds [23]. But the band gap is also affected by interactions between the transition metal ($d^{10}$) atoms, as discussed for the *RE*Cu*X*O (*X* = S, Se) compounds [86]. A comparable scenario was reported for the sulfide oxide LaCuSO, which is a transparent *p*-type semiconductor with a band gap around 3 eV [90]. Band structure calculations and photoemission experiments show consistently, that the valence band mainly consists of Cu states, hybridized with S-3*p* orbitals (Cu–S antibonding). The Cu-3*d* states are remarkably broadened probably due to direct Cu–Cu interactions and thus the band gap is strongly affected by the metal-metal distances. Of course, the gap also depends strongly on the energy of the lowest conduction band levels, which are the Cu-4*s* in LaCuSO and Zn-4*s* in *RE*ZnPO. Thus, the gap is mainly determined by the [CuS] or [ZnP] layers in LaCuSO and *RE*ZnPO, respectively. Consequently, a much smaller gap has been reported in BiCuOS (1.1 eV), where the conduction band is formed by the lower lying Bi-6*p* orbitals [44].

Metallic properties arise if the transition metal *d*-shell is partially occupied. In this case, the Fermi level is always dominated by the *d*-orbitals [89, 91], more or less hybridized with the *p*-orbitals of the particular main group element. According to the two-dimensional character, especially the $d_{x2-y2}$- and $d_{xy}$-orbitals of the metal atoms play a key role for the magnetic and presumably superconducting properties, because they generate flat bands leading to special Fermi surface features [89, 92]. The band filling puts these orbitals close to the Fermi level in the cases of LaFeAsO and LaFePO, but not in LaNiPO or LaNiAsO. This may be one reason for the fact that superconductivity occurs only at very low temperatures in the nickel compounds ($T_C < 4$ K) and could not be increased by doping. From this it is assumed that the pairing mechanisms of phosphide-oxide and arsenide-oxide superconductors are different [92].

The electronic structures and the influence of doping on the Fermi surface and magnetism of LaFeAsO and related compounds has been intensively studied by theoretical methods [91, 93]. The calculated magnetic moments are mostly too large (~1 $\mu_B$) in comparison with the experimental values (~0.35 $\mu_B$) [94], which indicate that the theoretical description of the new materials is not yet under control. Most results support the view that spin fluctuations play an important rule for the pairing, but the exact mechanism remains to be seen in the future. However, this field is rapidly growing and outside the scope of this review.

## Physical properties

### *Magnetic behavior*

Although a huge number of ZrCuSiAs type compounds are known, only few of them have been studied with respect to their magnetic properties. The basic magnetic data are summarized in Table 2. The investigated lanthanum compounds LaZnPO, LaZnAsO, LaZnSbO, LaCuSO, and LaCuTeO are all diamagnetic with only weak upturns of the susceptibilities at low temperatures, indicating only trace amounts of paramagnetic impurities. These data underline the closed-shell $d^{10}$ character of copper in the two chalcogenides. Different behavior has been observed for LaMnPO [33], where the manganese ions ($Mn^{2+}$ $d^5$) carry a magnetic moment and reveal ferromagnetic ordering at 320 K.

Susceptibility measurements of these diamagnets or Pauli paramagnets may be affected by ferromagnetic or antiferromagnetic impurity phases. Recent studies on Pauli paramagnetic LaFePO [87] revealed $Fe_2P$ ($T_C$ = 266 K [97]) and $FeSn_2$ ($T_N$ = 380 K [98]) as minor by products. Especially ferromagnetic $Fe_2P$ might influence the superconducting transition temperature (*vide infra*).

Several cerium based compounds have been studied in detail with respect to their magnetic behavior. CeZnAsO [64], CeZnSbO [64, 95], CeCuSO [40, 79, 83], and CeAgSO [40] are paramagnetic down to low temperatures without indications for magnetic ordering. Except CeCuSO, the experimentally determined magnetic moments are close to the free-ion value of 2.54 $\mu_B$ for $Ce^{3+}$. Takano et al. [83] recently reported that CeCuSO reveals a magnetic moment of only 1.62 $\mu_B$ per formula unit, much smaller than the free ion-value, indicating intermediate valent cerium. Specific heat measurements show a $\gamma$ value of about 324 mJ/molK$^2$, pointing to heavy Fermion character. This is also supported by resistivity measurements. Independent susceptibility measurements by Ueda *et al.* [79] and Chan *et al.* [40] revealed higher magnetic moments of 2.1 $\mu_B$ and furthermore, significant copper defects have been reported for $CeCu_{1-x}SO$ [39, 79] which certainly strongly influences the magnetic behavior. Further investigations on this system are necessary in order to understand the magnetic and electronic behavior in more detail.

The intermediate cerium valence of several of the cerium based compounds was already evident from the plots of the cell volumes [39, 49, 79]. In view of these negative deviations of the cerium cell volumes from the Iandelli plots it is interesting that the experimentally observed magnetic moment of CeFePO of 2.56 $\mu_B$ is even slightly higher than the free-ion value [48]. CeFePO is a magnetically non-ordered heavy Fermion metal with a Kondo temperature of 10 K and a $\gamma$value of 700 mJ/molK$^2$. In contrast, ferromagnetic ordering at $T_C$ = 15 K has been observed for CeRuPO [51], while LaRuPO is non-magnetic and behaves like a classical metal without anomalies down to 0.5 K. CeRuPO is a rare example of a ferromagnetic Kondo lattice system. The closely related compound CeOsPO orders antiferromagnetically at $T_N$ = 4.5 K and the magnetic ordering temperature decreases with increasing field strength.

Much higher ordering temperature of 85(4) K has been observed for SmCoPO, originally reported as 'SmCoP' [21]. This behavior has been ascribed to the ferromagnetic ordering of the cobalt magnetic moments. For the paramagnetic range a moment of 1.36(3) $\mu_B$/Co atom was deduced. UCuPO [13] has an even higher magnetic ordering temperature. The uranium magnetic moments show antiferromagnetic alignment below the Néel temperature of 220(2) K.

PrZnAsO, NdZnAsO [64], PrZnSbO [95], NdZnSbO [64], PrCuSO, and NdCuSO [79, 96] contain stable trivalent praseodymium and neodymium. The experimentally determined magnetic moments are in agreement with the corresponding free-ion values. All of these samples did not show magnetic ordering down to low temperatures (usually 4.2 K). The susceptibilities of PrCuSO and NdCuSO deviate from a simple Curie-Weiss law due to crystal field effects. The corresponding crystal field parameters have been determined by Nakao *et al.* [96] using an operator equivalent method.

The arsenide oxide LaFeAsO shows a spin-density-wave instability around 150 K which is associated with an abrupt structural distortion [54, 55]. The latter is associated with long-range antiferromagnetic ordering below ca. 134 K [55]. Evidence for the structural distortion was obtained from high-resolution neutron powder and synchrotron X-ray diffraction experiments. While de la Cruz *et al.* [55] suggest a symmetry reduction directly to the monoclinic system (space group *P*112/*n* but with constrained lattice parameters *a* and *b*), Nomura *et al.* [54] propose a symmetry reduction to the orthorhombic system, space group *Cmma*. On the basis of crystallographic group-subgroup relations [99-101] we consider the last symmetry reduction reduction as more probable since it is a simple *translationengleiche* transition of index 2 (t2) from *P*4/*nmm* to *Cmma* with a cell transformation *a*+*b*, *a*–*b*, *c*. This way the *a* and *b* parameters of the superstructure cell are decoupled and indeed one observes smaller *a* and larger *b* parameter in the ordered state [54]. The symmetry reduction to the monoclinic system requires at least two steps and the constraints on the lattice parameters [55] are questionable.

A very interesting situation occurs for the quaternary hydrides CeCoSiH [37], CeMnGeH, CeFeSiH, CeCoGeH [68, 102, 103], and CeRuSiH [70, 88], since these compounds have been obtained through hydrogenation of the ternary intermetallic silicides and germanides. This is in contrast to all other ZrCuSiAs type compounds, which are only stable as quaternary variants. Hydrogenation of the antiferromagnetic compounds CeCoSi and CeCoGe leads to spin fluctuation behavior in CeCoSiH and CeCoGeH [102, 103]. In the sequence CeMnGe → CeMnGeH the magnetic ordering of the cerium substructure is suppressed [71]. The temperature dependence of the magnetization of CeMnGeH is characteristic for a ferromagnetic, ferrimagnetic, or canted magnetic system. Two ordering temperatures, $T_{N1}$ = 313(2) and $T_{N2}$ = 41(2) K have been observed. In the sequence CeRuSi → CeRuSiH, hydrogenation changes the moderate heavy Fermion system to an antiferromagnet (the hydrogen insertion diminishes the influence of the Kondo effect) with two ordering temperatures, $T_{N1}$ = 7.5(2) and $T_{N2}$ = 3.1(2) K [70, 88]. The hydrogen content of LaCoGeH was proven by $^{1}$H MAS NMR [102].

*Superconductivity*

The first report on superconductivity in the ZrCuSiAs-type compound LaFePO was published in 2006 [25]. Even though it appeared remarkable to find superconductivity in an iron compound, the transition temperature ($T_C$) of 3.5 K was not spectacular at that time. The same holds true for LaNiPO, which becomes superconducting at 4.5 K [26, 27]. The transition temperature of phase pure LaFePO is probably higher because of the presence of ferromagnetic impurities in the samples (Fe, $Fe_2P$). Those can reduce $T_C$ due to the permanent magnetic field inside the sample. By treating finely grounded LaFePO with a magnet under liquid $N_2$, it is possible to remove most of those impurities and to increase $T_C$ from 3.5 to 7 K. [87].

The breakthrough came in February 2008, when *Kamihara* reported superconductivity at 26 K in $LaFeAs(O_{1-x}F_x)$ [28]. Then Chinese groups took over and even higher $T_C$ values followed quickly. By replacing the lanthanum ions with smaller rare earth ions, the $T_C$ increases strongly to 41 K in $CeFeAs(O_{1-x}F_x)$, 52 K in $PrFeAs(O_{1-x}F_x)$ and $NdFeAs(O_{1-x}F_x)$ and reached 55 K in $SmFeAs(O_{1-x}F_x)$ [104]. These transition temperatures can be regarded as lower limits, because of partially poor sample quality. $LaFeAs(O_{1-x}F_x)$ samples synthesized under high pressure [105] showed superconductivity at 41 K instead of 26 K. The authors ar-gued that the reason is a slightly smaller lattice parameter. However, the difference is very small and this seems unlikely to be the origin of this enormous increase of $T_C$. On the other side, the true fluoride content of the doped materials is unknown up to now. Most of the so far published *RE*FeAs($O_{1-x}F_x$) compounds contain significant amounts of impurity phases, among them *RE*OF, which makes the true content of fluoride inside the structure unreliable. This is supported by the finding, that fluoride-free, but oxygen deficient compounds *RE*FeAs$O_{1-x}$ become also superconducting between 30 and 55 K [61].

The undoped parent compound LaFeAsO is not superconducting, but shows distinctive anomalies of the physical properties around 150 K. The electrical resistivity and magnetic susceptibility drop at this temperature as a consequence of a structural phase transition [54, 55] (*vide ultra*). The latter has been assigned to the lock-in of a spin density wave (SDW) into long range antiferromagnetic ordering. The connection between the structural and magnetic transition was also proved by $^{57}$Fe-Mössbauer spectroscopy [106, 107] (*vide infra*).

The structural and magnetic phase transition of LaFeAsO is successively suppressed by increasing doping with electrons or holes. Doped materials like LaFeAs$O_{0.85}F_{0.15}$ (nominal) show no splitting of the $^{57}$Fe Mössbauer signal, thus the magnetic ordering is also suppressed. Superconductivity emerges by adding or removing 0.1-0.2 electrons per formula unit. It is generally believed that the nature of the SDW transition in undoped LaFeAsO is one major key to understand superconductivity in iron arsenides and the SDW seems to be mandatory for higher transitions temperatures. Indeed, this may be the reason for the low critical temperatures of the isostructural and isoelectronic phosphides *RE*FePO, where the SDW transition is not observed [87].

But even though several uncertainties, among them the role of the SDW transition as well as the exact chemical composition of the superconducting phases still exist, the new ZrCuSiAs-type superconductors have heralded a new age of superconductivity research. Bulk superconductivity at critical temperatures above 50 K has not been observed since the discovery of the cuprate superconductors by Bed-norz and Müller, more than twenty years ago [108]. Even if the pnictide oxide materials have not (yet) reached such high $T_C$, their discovery will undoubtedly give a fresh impetus to the great issue of superconductivity, namely the still open question about the mechanism of cooper pair formation at temperatures well above 40 K. Actually, more than twenty theories more or less related to the classical BCS theory from 1957, are not able to explain the coupling of the conduction electrons in these material conclusively [109].

Some properties of the ZrCuSiAs-type superconductors raise similarities to the cuprates. In both cases, non-superconducting antiferromagnetic parent structures exist. These are $La_2CuO_4$ in the case of the cuprates and LaFeAsO in the case of the pnictide oxides. Upon doping with holes or electrons, the antiferromagnetic state becomes unstable and superconductivity emerges. But there are also differences. $La_2CuO_4$ is an insulator, but LaFeAsO is a poor metal at room temperature. This is due to direct orbital interactions between the iron atoms at distances of 285 pm, whereas no direct *d*-orbital overlap occurs in the cuprates of the $La_2CuO_4$ family.

*Optical and opto-electronic properties*

Depending on the transition metal and group V / VI component, the various ZrCuSiAs type compounds show different color. The dimorphic (tetragonal ZrCuSiAs-type low-, and rhombohedral, NdZnPO-type high-temperature modification) *RE*ZnPO phosphide oxides are transparent with light yellow to dark red color [22, 23]. According to these relatively large band gaps, the model of ionic formula splitting discussed above is best applicable to these compounds. Single crystal absorption spectra measured for *RE*ZnPO (*RE* = Y, La, Pr, Nd, Sm, Dy) in the nir/vis region reveal unexpected variations for the optical band gap of these phosphide oxides. For *RE* = Pr, Nd, Sm, Dy, Ho *f*–*f* electronic transitions with nicely resolved ligand-field splittings are observed in the range 6000–20000 $cm^{-1}$. DFT band structure calculations show similarity between the valence bands of tetragonal and rhombohedral *RE*ZnPO as they possess mainly P-3*p* character. In both cases, the conduction bands have mainly Zn-4*s* character, but a significant contribution of *RE*-5*d* occurs in rhombohedral *RE*ZnPO, which is responsible for a smaller optical band gap for the latter compounds. Variations of the energy gaps of tetragonal *RE*ZnPO can be explained by hybridization of Zn-4*s* + *RE*-5*d* + *RE*-4*f* orbitals for the conduction band.

With the heavier group V and group VI elements, the band gaps become smaller and the compounds are black. Some of them are even metallic. To give an example, experimental data show optical band gaps of $E_g < 0.73$ and 0.71 eV for the black compounds $CeCu_{0.8}SO$ (*p*-type semiconductor) and CeAgSO at 298 K, respectively [40]. These values compare well with electronic structure calculations. The experimental optical band gaps for LaCuTeO and NdCuTeO are 2.31 and 2.26 eV, respectively [41]. According to positive values of the Seebeck and Hall coefficients, these compounds reveal *p*-type semiconductivity. Electronic structure calculations of these materials revealed, that the larger dispersion of the Cu 3*d* orbitals and the presence of Te 5*p* states near the valence band maximum (VBM) are responsible for the larger hole mobility of LaCuTeO as compared to LaCuSeO and LaCuSO.

The full ionic formula splitting is also valid for the fluoride sulfides. Grossholz and Schleid [45] observed smaragde green color for the platelet-shaped crystals of EuCuSF. Also SmCuSeF [74] shows deep color, while the fluoride BaZnPF [36] is brownish, indicating a smaller band gap. Upon doping *p*-type semiconductivity occurs for the solid solutions $Sr_{1-x}Na_xCuSF$ and $SrCuSO_xF_{1-x}$ [73].

Besides the recently discovered superconductivity in the pnictide oxides, the so far broadest investigations for ZrCuSiAs-type materials have been performed for the oxysulfides [12]. In their first report Ueda *et al.* [11] studied $La_{1-x}Sr_xCuSO$ (x = 0, 0.05) thin films prepared by radio-frequency sputtering [110] which showed high optical transmission of > 70% in the visible and near-infrared region and an energy gap of 3.1 eV. These materials are *p*-type semiconductors with a sharp photoluminescence peak at the optical absorption edge. The investigations have also been extended with respect to other diamagnetic rare earth elements, *i.e.* YCuSeO and LaCuSeO [86].

The widegap oxychalcogenides LaCu*Ch*O (*Ch* = chalcogen) are an outstanding family of materials, since widegap p-type conduction is rather difficult to realize. Furthermore, these materials exhibit high hole mobility, degenerate *p*-type conduction, room temperature exciton and large third order optical nonlinearity. All these highly interesting features have been summarized recently in a review by Hiramatsu *et al.* [12]. These authors also discuss a first application. It is possible to run a blue light-emitting diode at room temperature using an *pn* hetero-junction composed of a LaCuSO epilayer and an *n*-type amorphous $InGaZn_5O_8$ [12].

The concentration of lattice imperfections is an important parameter influencing the optical properties and the conduction behaviour. Takase *et al.* [111] have treated bars of LaCuSO under different thermal conditions, *i.e.* (a) 1070 K for 6 h, (b) 1170 K for 6 h, and (c) 1170 K for 40 h. An increase of lattice imperfections (evident also from the X-ray powder data) increases the intensity of the wide emission bands and the specific resistivity drastically decreases with increasing annealing time. Another important parameter on the resistivity behavior is doping. Koyano *et al.* [85] studied the whole series $(La_{1-x}Ca_x)Cu_{1-x}T_xSO$ (*T* = Mn, Co, Ni, Zn). Small changes in the interlayer distances shift the specific resistivity by several decades.

So far, ionic conductivity has only been reported for LaAgSO [8]. The single crystal X-ray data [9] revealed an enlarged displacement parameter for the silver atoms, although refinement of the occupancy parameters showed full occupancy. Electrochemical measurements with a cell $^{+}$Ag/LaAgSO/Pt$^{-}$ showed essentially ionic conductivity in the temperature regime 298 to 523 K with activation energy of 0.195 eV. Considering the defects on the transition metal sites of other copper and silver containing *RE*CuSO and *RE*AgSO compounds (*vide ultra*), one can also expect some copper and silver mobility in these materials.

*Mössbauer spectroscopy*

The three iron based compounds LaFePO [87], LaFeAsO, and $LaFeAsO_{0.89}F_{0.11}$ [106, 107] have been characterized by temperature dependent $^{57}$Fe Mössbauer spectroscopy. Measurements of LaFePO at 298, 77, 4.2 and 4 K show single sig-nals at isomer shifts around 0.35 mm/s, subject to weak quadrupole splitting [87]. At 4 K, a symmetric line broadening appears, resulting from a small transferred magnetic hyperfine field of 1.15(1) T and accompanied by an angle of 54.7(5) ° between $B_{hf}$ and $V_{zz}$, the main component of the electric field gradient tensor.

The different electronic states of iron in LaFeAsO and fluoride doped La-$FeAsO_{0.89}F_{0.11}$ have been studied in detail by $^{57}$Fe Mössbauer spectroscopy [106, 107]. The $^{57}$Fe spectra proved spin ordering in LaFeAsO and its suppression upon doping. The isomer shifts of the arsenide oxides are close to the data observed for the phosphide. Below the antiferromagnetic ordering ($T_N$ = 138 K) LaFeAsO shows full magnetic hyperfine field splitting with a hyperfine field of 4.86 T [106]. The magnetic moment at the iron atoms was estimated between 0.25–0.35 $\mu_B$/Fe atom.

Also the *RE*MnSbO (*RE* = La, Ce, Pr, Nd, Sm, Gd) and *RE*ZnSbO (*RE* = La, Ce, Pr) antimonide oxides have been studied via $^{121}$Sb Mössbauer spectroscopy [34]. In agreement with the crystal structure, all antimonides show single signals at isomer shifts ranging from –7.82 (LaMnSbO) to –8.37 mm/s (PrZnSbO). According to the ionic formula splitting discussed above the *RET*SbO compounds contains $Sb^{3-}$ pnictide anions. This Zintl anion also occur in the alkali metal antimonides $A_3Sb$ (A = Li, Na, K, Rb) with comparable isomer shifts of –7.3 mm/s for $Li_3Sb$ [112] and –8.39 mm/s for $Rb_3Sb$ [113]. Also the III–V semiconductors AlSb, GaSb and InSb [114] show comparable isomer shifts. Thus, the $^{121}$Sb Mössbauer spectroscopic data clearly underline the antimonide character.

**Conclusions and outlook**

So far, more than 150 representatives with the simple tetragonal ZrCuSiAs type structure (tP8) are known. These materials have intensively been investigated in the last 15 years with respect to their highly interesting physical properties: (i) the chalcogenides LaCu*Ch*O are widegap semiconductors with unique optoelectronic properties with substantial potential for LED application and optoelectronic devices, (ii) compounds like the ferromagnetic Kondo system CeRuPO are highly attractive materials for basic research, and (iii) several of the pnictide oxides have outstanding superconducting properties. In view of the rapidly growing work in this new research field (several new contributions are published daily on the preprint server (arXiv.org) of the Cornell University Library) we can expect further interesting results in the near future.

*Acknowledgements*

This work was financially supported by the Deutsche Forschungsgemeinschaft.

Table 1. Lattice parameters of the tetragonal compounds with ZrCuSiAs type structure.

| Compound | *a* (pm) | *c* (pm) | *V* ($nm^3$) | Reference |
|---|---|---|---|---|
| **phosphide oxides** | | | | |
| LaMnPO | 405.78 | 884.36 | 0.1456 | [33] |
| | 405.4(1) | 883.4(4) | 0.1452 | [18] |
| CeMnPO | 402.0(1) | 874.2(3) | 0.1413 | [18] |
| PrMnPO | 400.6(1) | 870.7(2) | 0.1397 | [18] |
| NdMnPO | 398.9(1) | 867.4(1) | 0.1380 | [18] |
| SmMnPO | 396.0(1) | 859.0(3) | 0.1347 | [18] |
| GdMnPO | 393.3(1) | 851.0(1) | 0.1316 | [18] |
| TbMnPO | 392.0(1) | 848.5(4) | 0.1304 | [18] |
| DyMnPO | 390.4(1) | 846.9(4) | 0.1291 | [18] |
| LaFePO | 396.4(1) | 851.2(3) | 0.1338 | [25] |
| | 395.70(9) | 850.7(4) | 0.1332 | [14] |
| | 396.10(1) | 851.58(2) | 0.1336 | [87] |
| | 396.2 | 851.1 | 0.1336 | [52] |
| | 396.307(4) | 850.87(1) | 0.1336 | [47] |
| CeFePO | 391.9(1) | 832.7(3) | 0.1279 | [14] |
| | 391.9(3) | 833.0(5) | 0.1279 | [48] |
| PrFePO | 391.13(6) | 834.5(2) | 0.1277 | [14] |
| NdFePO | 389.95(5) | 830.2(3) | 0.1262 | [14] |
| ’SmFeP’ | 388.03(6) | 819.8(4) | 0.1234 | [21] |
| SmFePO | 387.8(1) | 820.5(1) | 0.1234 | [14] |
| GdFePO | 386.1(3) | 812.3(7) | 0.1211 | [14] |
| LaCoPO | 396.78(9) | 837.9(3) | 0.1319 | [14] |
| | 396.81(9) | 837.79(1) | 0.1319 | [46] |
| CeCoPO | 392.13(7) | 821.9(4) | 0.1264 | [14] |
| ’PrCoP’ | 392.29(5) | 821.5(2) | 0.1264 | [21] |
| PrCoPO | 392.24(8) | 822.4(2) | 0.1265 | [14] |
| ’NdCoP’ | 390.44(9) | 817.3(3) | 0.1246 | [21] |
| NdCoPO | 390.84(5) | 817.2(2) | 0.1248 | [14] |
| ’SmCoP’ | 388.45(4) | 808.5(1) | 0.1220 | [21] |
| SmCoPO | 388.17(7) | 807.3(2) | 0.1216 | [14] |
| LaNiPO | 404.53(1) | 810.54(3) | 0.1326 | [27] |
| | 404.61(8) | 810.0(7) | 0.1326 | [26] |
| ThCuPO | 389.43(4) | 828.3(1) | 0.1256 | [15] |
| UCuPO | 379.3(1) | 823.3(2) | 0.1184 | [13] |
| LaZnPO | 404.0(1) | 890.8(2) | 0.1454 | [19] |
| | 402.759(14) | 887.105(12) | 0.1439 | [64] |
| | 404.109(1) | 890.486(3) | 0.1454 | [50] |
| α-CeZnPO | 401.3(1) | 882.4(2) | 0.1421 | [19] |
| | 401.08(9) | 882.0(4) | 0.1419 | [22] |
| α-PrZnPO | 399.3(2) | 877.2(7) | 0.1399 | [22] |
| ThAgPO | 396.6(1) | 878.6(3) | 0.1382 | [16, 49] |
| LaCdPO | 417.2(2) | 906.7(6) | 0.1578 | [78] |
| LaRuPO | 404.8(2) | 841.0(4) | 0.1378 | [51] |
| | 404.7(1) | 840.6(1) | 0.1377 | [14] |
| CeRuPO | 402.8(1) | 825.6(2) | 0.1340 | [51] |
| | 402.7(3) | 826(1) | 0.1340 | [32] |
| | 402.6(1) | 825.6(2) | 0.1338 | [14] |
| PrRuPO | 401.8(1) | 817.4(3) | 0.1320 | [14] |
| NdRuPO | 400.86(5) | 816.7(2) | 0.1312 | [14] |
| SmRuPO | 399.35(5) | 805.8(2) | 0.1285 | [14] |
| GdRuPO | 397.9(1) | 797.4(2) | 0.1262 | [14] |
| TbRuPO | 397.1(1) | 792.7(2) | 0.1250 | [16, 49] |
| DyRuPO | 396.25(1) | 787.2(2) | 0.1236 | [16, 49] |
| LaOsPO | 404.8(3) | 840.8(6) | 0.1378 | [16, 49] |
| CeOsPO | 402.8(1) | 828.7(3) | 0.1345 | [16, 49] |
| | 403.1(1) | 828.6(3) | 0.1346 | [51] |
| PrOsPO | 402.06(6) | 824.1(2) | 0.1332 | [16, 49] |
| NdOsPO | 401.01(1) | 819.2(2) | 0.1317 | [16, 49] |
| SmOsPO | 399.6(1) | 806.9(3) | 0.1288 | [16, 49] |
| **phosphide fluorides** | | | | |
| BaMnPF | 417.93(2) | 950.40(5) | 0.1660 | [35] |
| SrZnPF | 403.15(1) | 901.14(6) | 0.1465 | [35] |
| BaZnPF | 415.637(3) | 945.74(1) | 0.1634 | [36] |
| EuMnPF | 402.9(1) | 894.9(1) | 0.1453 | [66] |
| **phosphides and arsenides** | | | | |
| ZrCuSiP | 356.71(1) | 944.69(4) | 0.1202 | [67] |
| ZrCuSiAs | 367.36(2) | 957.12(9) | 0.1292 | [1] |
| HfCuSiAs | 363.4(1) | 960.1(1) | 0.1268 | [1] |
| **arsenide oxides** | | | | |
| YMnAsO | 395.7(1) | 875.0(6) | 0.1370 | [18] |
| LaMnAsO | 412.4(1) | 903.0(5) | 0.1536 | [18] |
| CeMnAsO | 408.6(1) | 895.6(2) | 0.1495 | [18] |
| PrMnAsO | 406.7(1) | 891.9(3) | 0.1475 | [18] |
| NdMnAsO | 404.9(2) | 889.3(1) | 0.1458 | [18] |
| SmMnAsO | 402.0(1) | 882.9(3) | 0.1427 | [18] |
| GdMnAsO | 398.9(1) | 880.5(3) | 0.1401 | [18] |
| TbMnAsO | 397.8(1) | 874.3(4) | 0.1384 | [18] |
| DyMnAsO | 395.9(1) | 872.7(4) | 0.1368 | [18] |
| UMnAsO | 386.9(1) | 852.5(2) | 0.1276 | [18] |
| LaFeAsO | 403.552(8) | 873.93(2) | 0.1423 | [28] |
| | 403.2 | 872.6 | 0.1419 | [62] |
| | 403.007(9) | 873.68(2) | 0.1419 | [55] |
| | 403.397(4) | 874.49(4) | 0.1423 | [63] |
| | 403.8(1) | 875.3(6) | 0.1427 | [20] |
| | 403.268(1) | 874.111(4) | 0.1422 | [54] |
| $LaFeAsO_{1-x}F_x$ | 402.4 | 871.7 | 0.1412 | [57] |
| | 403.0(1) | 870.6(2) | 0.1414 | [58] |
| CeFeAsO | 400.0(1) | 865.5(1) | 0.1385 | [20] |
| | 399.6 | 864.8 | 0.1381 | [59] |
| PrFeAsO | 398.5(1) | 859.5(3) | 0.1365 | [20] |
| $PrFeAsO_{1-x}F_x$ | 396.7(1) | 856.1(3) | 0.1347 | [29] |
| NdFeAsO | 396.5(1) | 857.5(2) | 0.1348 | [20] |
| SmFeAsO | 394.0(1) | 849.6(3) | 0.1319 | [20] |
| | 393.3(5) | 849.5(4) | 0.1314 | [61] |
| | 394.0 | 849.6 | 0.1319 | [56] |
| GdFeAsO | 391.5(1) | 843.5(4) | 0.1293 | [20] |
| $GdFeAsO_{1-x}F_x$ | 400.1 | 865.0 | 0.1385 | [60] |
| LaCoAsO | 405.4(1) | 847.2(3) | 0.1392 | [20] |
| | 405.26(1) | 846.20(4) | 0.1390 | [46] |
| CeCoAsO | 401.5(1) | 836.4(2) | 0.1348 | [20] |
| PrCoAsO | 400.5(1) | 834.4(2) | 0.1338 | [20] |
| NdCoAsO | 398.2(1) | 831.7(4) | 0.1319 | [20] |
| LaNiAsO | 412.309(1) | 818.848(6) | 0.1392 | [53] |
| ThCuAsO | 396.14(5) | 844.0(1) | 0.1324 | [15] |
| YZnAsO | 394.3(1) | 884.3(3) | 0.1375 | [19] |
| LaZnAsO | 409.5(1) | 906.8(3) | 0.1521 | [19] |
| | 410.492(13) | 908.178(47) | 0.1530 | [64] |
| CeZnAsO | 406.9(1) | 899.5(3) | 0.1489 | [19] |
| | 406.79(5) | 899.8(1) | 0.1489 | [64] |
| PrZnAsO | 404.7(1) | 896.3(1) | 0.1468 | [19] |
| | 404.77(2) | 896.6(2) | 0.1469 | [64] |
| NdZnAsO | 403.0(1) | 894.9(4) | 0.1453 | [19] |
| | 402.95(6) | 895.2(1) | 0.1454 | [64] |
| SmZnAsO | 400.3(1) | 890.3(2) | 0.1427 | [19] |
| GdZnAsO | 397.6(1) | 889.4(3) | 0.1406 | [19] |
| TbZnAsO | 395.7(1) | 884.1(2) | 0.1384 | [19] |
| DyZnAsO | 394.7(1) | 883.8(1) | 0.1377 | [19] |
| LaCdAsO | 412.9(2) | 923.0(3) | 0.1643 | [78] |
| CeCdAsO | 419.1(1) | 917.1(4) | 0.1611 | [78] |
| PrCdAsO | 417.2(1) | 913.6(4) | 0.1590 | [78] |
| NdCdAsO | 415.1(1) | 912.3(5) | 0.1572 | [78] |
| (Nd, Sm)CdAsO | 414.7(2) | 912.5(6) | 0.1569 | [78] |
| LaRuAsO | 411.9(1) | 848.8(1) | 0.1440 | [20] |
| CeRuAsO | 409.6(1) | 838.0(3) | 0.1406 | [20] |
| PrRuAsO | 408.5(1) | 833.7(1) | 0.1391 | [20] |
| NdRuAsO | 407.9(1) | 829.2(2) | 0.1380 | [20] |
| SmRuAsO | 405.0(2) | 819.1(7) | 0.1343 | [20] |
| GdRuAsO | 403.9(1) | 811.8(6) | 0.1324 | [20] |
| TbRuAsO | 402.7(1) | 807.8(1) | 0.1310 | [20] |
| DyRuAsO | 402.2(2) | 805.0(3) | 0.1302 | [20] |
| **antimonide oxides** | | | | |
| LaMnSbO | 423.95(7) | 955.5(2) | 0.1717 | [34] |
| | 424.2(1) | 955.7(2) | 0.1720 | [18] |
| CeMnSbO | 420.8(1) | 950.7(1) | 0.1683 | [34] |
| | 421.8(1) | 951.7(2) | 0.1693 | [18] |
| PrMnSbO | 418.8(1) | 947.2(3) | 0.1661 | [34] |
| | 418.7(1) | 946.0(1) | 0.1658 | [18] |
| NdMnSbO | 416.6(1) | 947.1(3) | 0.1644 | [34] |
| | 416.5(1) | 946.2(2) | 0.1641 | [18] |
| SmMnSbO | 413.1(1) | 942.3(1) | 0.1608 | [34] |
| | 413.5(1) | 941.8(2) | 0.1610 | [18] |
| GdMnSbO | 410.0(1) | 942.1(2) | 0.1584 | [34] |
| | 409.0(1) | 941.0(1) | 0.1574 | [18] |
| TbMnSbO | 408.3(1) | 939.2(6) | 0.1566 | [34] |
| LaZnSbO | 422.67(6) | 953.8(2) | 0.1704 | [34] |
| | 422.62(2) | 953.77(6) | 0.1704 | [17] |
| | 422.604(7) | 953.691(24) | 0.1703 | [64] |
| CeZnSbO | 419.9(1) | 948.7(2) | 0.1673 | [34] |
| | 419.76(4) | 947.4(1) | 0.1669 | [17] |
| | 419.66(2) | 947.96(4) | 0.1669 | [64] |
| PrZnSbO | 418.79(8) | 946.7(5) | 0.1660 | [34] |
| | 417.63(4) | 945.1(1) | 0.1648 | [17] |
| NdZnSbO | 415.9(1) | 945.4(4) | 0.1635 | [34] |
| | 415.81(2) | 944.95(5) | 0.1634 | [17] |
| | 415.78(2) | 944.33(5) | 0.1632 | [64] |
| SmZnSbO | 412.80(2) | 940.16(6) | 0.1602 | [17] |
| **antimonide fluorides** | | | | |
| BaZnSbF | 443.84(2) | 977.89(6) | 0.1926 | [35] |
| **bismuthides** | | | | |
| $LaNiBiO_{0.8}$ | 407.3 | 930.1 | 0.1543 | [65] |
| **sulphide oxides** | | | | |
| BiCuSO | 387.05(5) | 856.1(1) | 0.1283 | [82] |
| | 387.08(8) | 855.8(1) | 0.1282 | [43] |
| | 386.91(1) | 856.02(4) | 0.1281 | [44] |
| LaCuSO | 399.9(1) | 853(4) | 0.1364 | [10] |
| | 399.484(2) | 851.254(5) | 0.1358 | [85] |
| | 399.5 | 851.0 | 0.1358 | [42] |
| | 399.6 | 851.7 | 0.1360 | [86] |
| | 399.38(2) | 852.15(4) | 0.1359 | [44] |
| | 399.625(4) | 851.743(9) | 0.1360 | [79] |
| CeCuSO | 392.3 | 833.3 | 0.1282 | [83] |
| | 392.28(1) | 834.75(3) | 0.1285 | [79] |
| $CeCu_{1-x}SO$ | 391.9(1) | 843.2(3) | 0.1295 | [39] |
| $CeCu_{0.95}SO$ | 391.7(2) | 841.2(3) | 0.1290 | [39] |
| $CeCu_{0.90}SO$ | 391.8(1) | 839.4(5) | 0.1288 | [39] |
| $CeCu_{0.85}SO$ | 391.1(1) | 836.5(3) | 0.1279 | [39] |
| $CeCu_{0.809}SO$ | 391.42(3) | 829.80(10) | 0.1271 | [40] |
| $CeCu_{0.762}SO$ | 390.72(3) | 828.34(10) | 0.1265 | [40] |
| PrCuSO | 3.9419(4) | 843.98(9) | 0.1311 | [84] |
| | 393.9(1) | 843.8(3) | 0.1309 | [39] |
| | 394.1 | 843.8 | 0.1311 | [86] |
| | 394.148(9) | 843.76(1) | 0.1311 | [79] |
| NdCuSO | 390.3(5) | 848.0(1) | 0.1292 | [81] |
| | 391.8(1) | 843.4(6) | 0.1295 | [39] |
| | 392 | 842.8 | 0.1295 | [86] |
| | 391.96(1) | 842.82(2) | 0.1295 | [79] |
| SmCuSO | 385.4(3) | 844.2(7) | 0.1254 | [81] |
| | 388.7(3) | 838.5(7) | 0.1267 | [39] |
| EuCuSO | 387.4(2) | 837.9(6) | 0.1258 | [39] |
| LaAgSO | 405.0(2) | 903.9(3) | 0.1483 | [9] |
| | 406.6(1) | 909.5(1) | 0.1504 | [8] |
| $CeAg_{0.777}SO$ | 392.56(3) | 899.43(9) | 0.1386 | [40] |
| $CeAg_{0.763}SO$ | 392.30(3) | 901.30(8) | 0.1387 | [40] |
| **sulphide fluorides** | | | | |
| SrCuFS | 395.70(2) | 865.98(7) | 0.1356 | [73] |
| BaCuSF | 412.30(1) | 903.27(2) | 0.1535 | [72] |
| EuCuSF | 394.74(3) | 864.25(6) | 0.1347 | [45] |
| **selenide oxides** | | | | |
| BiCuSeO | 392.13(1) | 891.33(5) | 0.1371 | [82] |
| | 392.87(1) | 892.91(2) | 0.1378 | [44] |
| YCuSeO | 389.89(1) | 863.20(2) | 0.1312 | [38] |
| | 387.89(1) | 873.11(5) | 0.1314 | [86] |
| | 388.09(1) | 874.34(1) | 0.1317 | [77] |
| LaCuSeO | 408.98(2) | 893.75(3) | 0.1495 | [38] |
| | 406.5(1) | 879.2(1) | 0.1453 | [39] |
| | 406.7 | 879.8 | 0.1455 | [86] |
| | 406.70(1) | 880.06(8) | 0.1456 | [44] |
| | 406.65(3) | 881.19(2) | 0.1457 | [77] |
| CeCuSeO | 401.85(15) | 871.14(14) | 0.1407 | [77] |
| $CeCu_{1-x}SeO$ | 398.8(1) | 875.0(4) | 0.1391 | [39] |
| $CeCu_{0.95}SeO$ | 398.8(1) | 868.7(3) | 0.1382 | [39] |
| $CeCu_{0.90}SeO$ | 398.3(2) | 868.6(5) | 0.1377 | [39] |
| $CeCu_{0.85}SeO$ | 397.7(1) | 866.7(3) | 0.1370 | [39] |
| $CeCu_{0.80}SeO$ | 397.9(2) | 864.9(4) | 0.1369 | [39] |
| PrCuSeO | 400.4(3) | 873.3(4) | 0.1400 | [39] |
| NdCuSeO | 398.6(1) | 883.3(4) | 0.1403 | [82] |
| | 398.54(4) | 874.67(2) | 0.1389 | [77] |
| SmCuSeO | 395.12(1) | 871.57(2) | 0.1361 | [38] |
| | 395.63(1) | 871.64(1) | 0.1364 | [77] |
| EuCuSeO | 393.6(1) | 871.6(4) | 0.1350 | [39] |
| GdCuSeO | 392.73(1) | 867.72(2) | 0.1338 | [38] |
| | 391.86(2) | 872.3(5) | 0.1339 | [76] |
| | 392.06(8) | 874.3(2) | 0.1344 | [82] |
| | 392.08(5) | 870.68(1) | 0.1338 | [77] |
| TbCuSeO | 390.2(1) | 870.5(2) | 0.1325 | [39] |
| | 390.58(4) | 871.71(1) | 0.1330 | [77] |
| DyCuSeO | 388.54(2) | 870.1(4) | 0.1314 | [76] |
| | 388.65(6) | 870.9(1) | 0.1316 | [82] |
| | 389.57(2) | 875.29(2) | 0.1328 | [77] |
| HoCuSeO | 386.6(2) | 868.5(7) | 0.1298 | [76] |
| | 387.0(1) | 871.1(2) | 0.1304 | [39] |
| | 387.25(3) | 870.28(1) | 0.1305 | [77] |
| ErCuSeO | 385.6(1) | 871.1(2) | 0.1295 | [39] |
| LaAgSeO | 412.5(1) | 930.0(2) | 0.1582 | [80] |
| CeAgSeO | 400.9(2) | 934.7(5) | 0.1502 | [80] |
| PrAgSeO | 406.9(1) | 929.3(4) | 0.1539 | [80] |
| NdAgSeO | 404.1(1) | 931.3(4) | 0.1521 | [80] |
| $Nd_{0.6}Sm_{0.4}AgSeO$ | 402.8(3) | 930.1(11) | 0.1509 | [80] |
| **telluride oxides** | | | | |
| BiCuTeO | 404.11(2) | 952.37(5) | 0.1555 | [44] |
| LaCuTeO | 417.92(1) | 933.17(3) | 0.1630 | [81] |
| | 418.15(3) | 934.17(7) | 0.1633 | [75] |
| | 417.75(5) | 932.60(16) | 0.1628 | [41] |
| | 418.08(2) | 934.41(8) | 0.1633 | [44] |
| CeCuTeO | 414.97(3) | 930.90(10) | 0.1603 | [41] |
| | 409.7(1) | 930.3(3) | 0.1562 | [39] |
| PrCuTeO | 412.1(1) | 929.9(5) | 0.1578 | [39] |
| NdCuTeO | 410.56(9) | 933.2(4) | 0.1573 | [41] |
| | 409.7(1) | 931.8(4) | 0.1564 | [39] |
| LaAgTeO | 423.3(2) | 984.1(5) | 0.1763 | [80] |
| **selenide fluorides** | | | | |
| BaCuSeF | 423.91(1) | 912.17(2) | 0.1639 | [72] |
| SmCuSeF | 405.81(3) | 881.32(7) | 0.1451 | [74] |
| **hydrides** | | | | |
| CeMnSiH | 407.7 | 791.7 | 0.1316 | [69] |
| CeMnGeH | 414.1 | 794.0 | 0.1362 | [69] |
| | 414.08(2) | 794.01(3) | 0.1361 | [71] |
| CeFeSiH | 399.4(2) | 781.0(4) | 0.1246 | [71] |
| CeCoSiH | 395.5(2) | 786.1(3) | 0.1230 | [37] |
| LaCoGeH | 406.3(2) | 791.2(3) | 0.1306 | [68] |
| CeCoGeH | 404.0(2) | 773.5(4) | 0.1262 | [68] |
| LaRuSiH | 420.2(1) | 772.1(1) | 0.1363 | [70] |
| CeRuSiH | 417.98(5) | 751.20(7) | 0.1312 | [70] |
| CeRuSiH | 417.77(1) | 750.72(1) | 0.1302 | [88] |

Table 2. Magnetic properties of various ZrCuSiAs type compounds. D: diamagnetism; P: paramagnet, AF: antiferromagnet, F: ferromagnet, SF: spin fluctuation; IV: intermediate valence; $T_N$: Néel temperature, $T_C$: Curie temperature, $\mu_{exp}$: experimental magnetic moment, Θ: paramagnetic Curie temperature (Weiss constant).

| Compound | magnetism | $\mu_{exp}$ / $\mu_B$ | $T_N$, $T_C$ / K | Θ / K | Ref. |
|---|---|---|---|---|---|
| **pnictides** | | | | | |
| LaMnPO | F | – | 320 | – | [33] |
| LaZnPO | D | – | – | – | [64] |
| CeFePO | P | 2.56 | – | –52 | [48] |
| LaCoPO | F | – | 43 | – | [46] |
| ’SmCoP’ | F | 1.36(3) | 85(4) | 115(3) | [21] |
| UCuPO | AF | 2.68 | 220 | – | [13] |
| CeRuPO | F | 2.3 | 15 | 8 | [51] |
| CeOsPO | AF | 2.45 | 4.5 | –9 | [51] |
| LaCoAsO | F | – | 66 | – | [46] |
| LaZnAsO | D | – | – | – | [64] |
| CeZnAsO | P | 2.52 | – | – | [64] |
| PrZnAsO | P | 3.58 | – | – | [64] |
| NdZnAsO | P | 3.45 | – | – | [64] |
| LaZnSbO | D | – | – | – | [64] |
| CeZnSbO | P | 2.43 | – | – | [64] |
| CeZnSbO | P | 2.37 | – | –14.1 | [95] |
| PrZnSbO | P | 3.28 | – | –18.3 | [95] |
| NdZnSbO | P | 3.33 | – | – | [64] |
| **chalcogenides** | | | | | |
| LaCuSO | D | – | – | – | [96] |
| CeCuSO | IV | 1.62 | – | –5.3 | [83] |
| | P | 2.1 | – | – | [79] |
| $CeCu_{0.8}SO$ | P | 2.13(6) | – | –30.2(1) | [40] |
| PrCuSO | P | 3.5 | – | – | [96] |
| | P | 3.1 | – | – | [79] |
| NdCuSO | P | 3.7 | – | – | [96] |
| | P | 3.3 | – | – | [79] |
| CeAgSO | P | 2.10(1) | – | –28.1(3) | [40] |
| LaCuTeO | D | – | – | – | [75] |
| **hydrides** | | | | | |
| CeCoSiH | SF | 2.72(5) | – | –56(1) | [37] |
| CeCoGeH | SF | 2.78(5) | – | –29(1) | [68] |
| CeRuSiH | AF | 2.59(3) | 7.5(2); 3.1(2) | –18 | [70] |

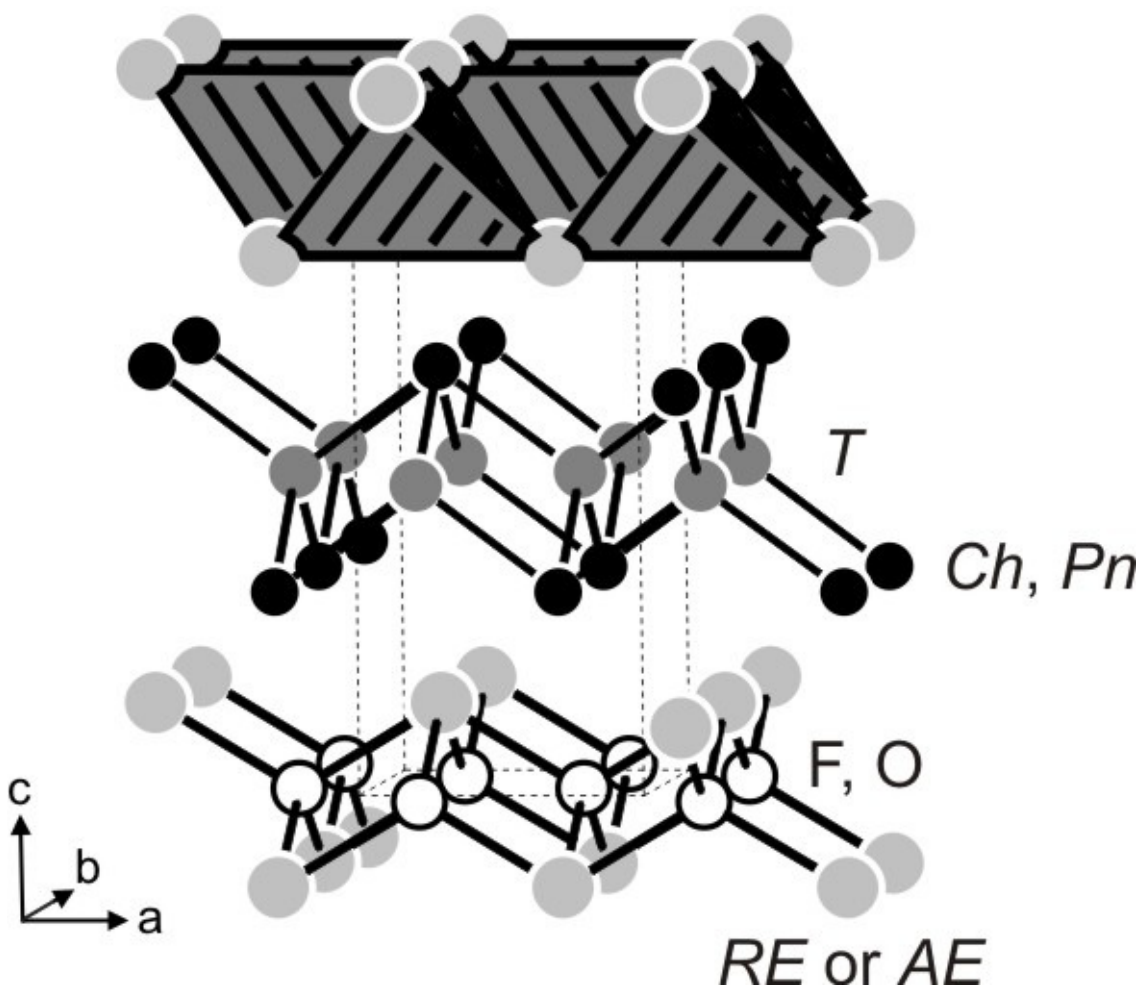

Fig. 1. The crystal structure of the tetragonal ZrCuSiAs type compounds. The different layers of condensed tetrahedra are emphasized. For details see text.